\documentclass[journal,a4paper]{IEEEtran}

\usepackage{array}
\usepackage{amsmath}
\usepackage{amssymb}
\usepackage{upref}
\usepackage{amsfonts}
\usepackage{graphicx}
\usepackage{cite}  
\usepackage{wasysym}

\usepackage{mathrsfs}

\newcommand{\field}[1]{\mathbb{#1}}

\newcommand{\F}{\field{F}}
\newcommand{\Z}{\field{Z}}

\newcommand{\cF}{{\cal F}}

\newcommand{\cG}{{\cal G}}

\newcommand{\cP}{{\cal P}}

\newcommand{\sP}{\cP}
\newcommand{\sG}{\cG}

\newcommand{\Gr}{\smash{{\sG\kern-1.5pt}_q\kern-0.5pt(n,k)}}
\newcommand{\Gk}{\smash{{\sG\kern-1.5pt}_q\kern-0.5pt(n,k_1)}}
\newcommand{\Gkk}{\smash{{\sG\kern-1.5pt}_q\kern-0.5pt(n,k_2)}}
\newcommand{\Grtwo}{\smash{{\sG\kern-1.5pt}_2\kern-0.5pt(n,k)}}
\newcommand{\Gkone}{\smash{{\sG\kern-1.5pt}_q\kern-0.5pt(n,k_1)}}
\newcommand{\Gktwo}{\smash{{\sG\kern-1.5pt}_q\kern-0.5pt(n,k_2)}}
\newcommand{\Ps}{\smash{{\sP\kern-2.0pt}_q\kern-0.5pt(n)}}

\newtheorem{theorem}{Theorem}

\newtheorem{lemma}{Lemma}
\newtheorem{remark}{Remark}

\newtheorem{example}{Example}

\newenvironment{algorithm}{%
       \begin{minipage}{\columnwidth}\vspace{0.5ex}%
       \makebox[0ex]{}\hrulefill\makebox[0ex]{}\\*\normalsize}{%
             \makebox[0ex]{}\hrulefill\makebox[0ex]{}\end{minipage}}

\begin{document}

\bibliographystyle{IEEEtran}

\title{Enumerative Coding for Grassmannian Space}
\author{Natalia Silberstein and Tuvi
Etzion,~\IEEEmembership{Fellow,~IEEE}
\thanks{N. Silberstein is with the Department of Computer Science,
Technion --- Israel Institute of Technology, Haifa 32000, Israel.
(email: natalys@cs.technion.ac.il). This work is part of her Ph.D.
thesis performed at the Technion.}
\thanks{T. Etzion is with the Department of Computer Science,
Technion --- Israel Institute of Technology, Haifa 32000, Israel.
(email: etzion@cs.technion.ac.il).}
\thanks{The material in this paper was presented in part in the 2009 IEEE
Information Theory Workshop, Taormina, Sicily, Italy, October
2009.}
\thanks{This work was supported in part by the Israel
Science Foundation (ISF), Jerusalem, Israel, under Grant 230/08.}
}

\maketitle

\begin{abstract}
The Grassmannian space $\Gr$ is the set of all $k-$dimensional
subspaces of the vector space~\smash{$\F_q^n$}. Recently, codes in
the Grassmannian have found an application in network coding. The
main goal of this paper is to present efficient enumerative
encoding and decoding techniques for the Grassmannian. These
coding techniques are based on two different orders for the
Grassmannian induced by different representations of
$k$-dimensional subspaces of $\F_q^n$. One enumerative coding
method is based on a Ferrers diagram representation and on an
order for $\Gr$ based on this representation. The complexity of
this enumerative coding is $O(k^{5/2} (n-k)^{5/2})$ digit
operations. Another order of the Grassmannian is based on a
combination of an identifying vector and a reduced row echelon
form representation of subspaces. The complexity of the
enumerative coding, based on this order, is $O(nk(n-k)\log
n\log\log n)$ digits operations. A combination of the two methods
reduces the complexity on average by a constant factor.
\end{abstract}


\section{Introduction}
\label{sec:introduction} Let $\F_q$ be a finite field of size $q$.
The\textit{ Grassmannian space} (Grassmannian, in short), denoted
by $\Gr$, is the set of all $k$-dimensional subspaces of the
vector space~\smash{$\F_q^n$}, for any given two integers $k$ and
$n$, $0 \le k \le n$. It is well known~\cite{vLWi92} that $| \Gr |
=\begin{footnotesize}\left[\begin{array}{c}n\\k\end{array}\right]_{q}\end{footnotesize}$,
where
$\begin{footnotesize}\left[\begin{array}{c}n\\k\end{array}\right]_{q}\end{footnotesize}$
is a $q$-ary Gaussian coefficient, defined by

\begin{equation}
\label{eq:num_subspace}
\begin{small}\left[\begin{array}{c}n\\k\end{array}\right]\end{small}_{q}=\prod_{i=0}^
{k-1}\frac{q^{n-i}-1} {q^{k-i}-1}~,
\end{equation}
where
$\begin{footnotesize}\left[\begin{array}{c}n\\0\end{array}\right]_{q}\end{footnotesize}=1$,
and
$\begin{footnotesize}\left[\begin{array}{c}n\\k\end{array}\right]_{q}\end{footnotesize}=0$
if $k> n$ or $k < 0$.

\vspace{0.05cm}

Coding (and related designs) in the Grassmannian was considered in
the last forty years,
e.g.~\cite{Knu71,Tho87,MaZh95,Tho96,AAK,ScEt02,BKL05}. Koetter and
Kschischang~\cite{KK} presented an application of error-correcting
codes in  $\Gr$ to random network coding. This application has
motivated extensive work in the
area~\cite{XiFu07,EV08,MGR08,SKK08,SiKs09,GaYa08,GaYa08A,GaYa09,EtSi09,KoKu08,Ska08}.
A natural question is how to encode/decode the subspaces in the
Grassmannian in an efficient way. By encoding we mean a
transformation of an information word into a $k$-dimensional
subspace. Decoding is the inverse transformation of the
$k$-dimensional subspace into the information word.

To solve this coding problem, we will use the general enumerative
coding method which was presented by Cover~\cite{Cover}. Let
$\{0,1\}^n$ denote the set of all binary vectors of length $n$.
Let $S$ be a subset of $\{0,1\}^n$. Denote by
$n_S(x_1,x_2,\ldots,x_k)$ the number of elements of $S$ for which
the first $k$ coordinates are given by $(x_1,x_2,\ldots,x_k)$,
where $x_1$ is the most significant bit. A lexicographic order of
$S$ is defined as follows. We say that for $x,y\in \{0,1\}^n$,
$x<y$, if $x_k<y_k$ for the least index $k$ such that $x_k\neq
y_k$. For example, $00101<00110$.
\begin{theorem}\cite{Cover}
\label{thm:cover} The lexicographic index (decoding) of $x\in~S$
is given by
$$\text{ind}_S(x)=\sum_{j=1}^{n}x_j \cdot n_S(x_1,x_2,\ldots,x_{j-1},0).$$
\end{theorem}

Let $S$ be a given subset and let $i$ be a given index. The
following algorithm finds the unique element $x$ of the subset $S$
such that $\text{ind}_S(x)=i$ (encoding).

\textit{Inverse algorithm}~\cite{Cover}:
For $k=1,\ldots,n$, if $i\geq n_S(x_1,x_2,\ldots,x_{k-1},0)$ then
set $x_k=1$ and $i=i-n_S(x_1,x_2,\ldots,x_{k-1},0)$; otherwise set
$x_k=0$.

\begin{remark}
The coding algorithms of Cover are efficient if
$n_S(x_1,x_2,\ldots,x_{j-1},0)$ can be calculated efficiently.
\end{remark}

Cover~\cite{Cover} also presented the extension of these results
to arbitrary finite alphabets. For our purpose this extension is
more relevant as we will see in the sequel. The formula for
calculating the lexicographic index of $x\in
S\subseteq\{1,2,3,\ldots,M\}^n$ is given as follows.
\begin{equation}\text{ind}_S(x)=\sum_{j=1}^{n}\sum_{m<x_j}n_S(x_1,x_2,\ldots,x_{j-1},m).\label{cover}
\end{equation}



Enumerative coding has various applications and it was considered
in many papers, e.g.~\cite{BrIm00,Kur02,Imm99}. Our goal in this
paper is to apply this scheme to the set of all subspaces in a
Grassmannian, using different lexicographic orders. These
lexicographic orders are based on different representations of
subspaces. Lexicographic orders also have other applications, e.g.
in constructions of lexicographic codes (lexicodes)~\cite{CoSl86}.

The rest of this paper is organized as follows. In
Section~\ref{sec:form} we discuss different representations of
subspaces in the Grassmannian. We define the reduced row echelon
form of a $k$-dimensional subspace and its Ferrers diagram. These
two concepts combined with the identifying vector of a
subspace~\cite{EtSi09} will be our main tools for the
representation of subspaces. We also define and discuss some type
of partitions which have an important role in our exposition. In
Section~\ref{sec:Cover} we present a new lexicographic order for
the Grassmannian based on a representation of a subspace by its
identifying vector and its reduced row echelon form. For this
order we describe an enumerative coding method, whose computation
complexity is $O(nk(n-k)\log n\log\log n)$ digit operations per
subspace. In Section~\ref{sec:Ferrers} we discuss the more
intuitive order for the Grassmannian based on Ferrers diagram
representation and present a second enumerative coding method for
the Grassmannian. In Section~\ref{sec:combination} we show how  we
can combine the two coding methods mentioned above to find a more
efficient enumerative coding for the Grassmannian. In
Section~\ref{sec:conclude} we summarize our results and discuss
some related problems.

\section{Representation of Subspaces and Partitions}
\label{sec:form}

In this section we give the definitions for two concepts which are
useful in describing a subspace in $\Gr$: Ferrers diagram (which
is defined in connection to a partition) and reduced row echelon
form. Based on these concepts we present two representations for
subspaces from which our enumerative coding techniques will be
induced. Representation of subspaces is also important in other
problems related to the Grassmannian. For example, in constructing
error-correcting codes in the Grassmannian~\cite{EtSi09,SiEt10}.

A \textit{partition} of a positive integer $m$ is a representation
of $m$ as  a sum of positive integers, not necessarily distinct.
We order this collection of integers in a decreasing order. The
partition function $p(m)$ is the number of different partitions of
$m$~\cite{vLWi92,And84,Sta86}.


A {\it Ferrers diagram} $\cF$ represents a partition as a pattern
of dots with the $i$-th row having the same number of dots as the
$i$-th term in the partition~\cite{vLWi92,And84,Sta86} (In the
sequel, a {\it dot} will be denoted by a $"\bullet"$). A Ferrers
diagram satisfies the following conditions.
\begin{itemize}
\item The number of dots in a row is at most the number of dots in
the previous row.

\item All the dots are shifted to the right of the diagram.
\end{itemize}

\begin{remark} Our definition of Ferrers diagram (see~\cite{EtSi09})
is slightly different form the usual
definition~\cite{vLWi92,And84,Sta86}, where the dots in each row
are shifted to the left of the diagram.
\end{remark}

A $k$-dimensional subspace $X\in \Gr$ can be represented by a
$k\times n$ matrix, whose rows form a basis for $X$. Such a $k
\times n$ matrix is in {\it reduced row echelon form} (RREF in
short) if the following conditions are satisfied.
\begin{itemize}
\item The leading coefficient (pivot) of a row is always to the
right of the leading coefficient of the previous row.

\item All leading coefficients are {\it ones}.

\item Every leading coefficient is the only nonzero entry in its
column.
\end{itemize}
For a given subspace $X$, there is exactly one matrix in RREF and
it will be denoted by $\mbox{RE} (X)$. For simplicity, we will
assume that the entries in $\mbox{RE} (X)$ are taken from $\Z_q$
instead of $\F_q$, using an appropriate bijection.

The {\it Ferrers tableaux form} of a subspace $X$, denoted by
$\cF(X)$, is obtained by removing from each row of $\mbox{RE}(X)$
the leading coefficient and the {\it zeroes} to the left of it.
All the remaining entries are shifted to the right. $\cF(X)$
defines a unique representation of $X$. The Ferrers diagram of
$X$, denoted by $\cF_X$, is obtained from $\cF(X)$ by replacing
the entries of $\cF(X)$ with dots.

\vspace{0.1cm}

\begin{example}
\label{exm:running} We consider a 3-dimensional subspace $X$ of
$\F_2^7$ with the following $3 \times 7$ matrix in RREF given by

\begin{footnotesize}
\begin{align*}
\mbox{RE}(X)=\left( \begin{array}{ccccccc}
1 & {\bf 0} & 0 & 0 & {\bf 1} & {\bf 1} & {\bf 0} \\
0 & 0 & 1 & 0 & {\bf 1} & {\bf 0} & {\bf 1} \\
0 & 0 & 0 & 1 & {\bf 0} & {\bf 1} & {\bf 1}
\end{array}
\right) .
\end{align*}
\end{footnotesize}\\
Its Ferrers tableaux form and Ferrers diagram are given by
\begin{align*}
\begin{footnotesize}
\cF(X)=
\begin{array}{cccc}
0 & 1 & 1 & 0 \\
&1 & 0 & 1  \\
&0 & 1 & 1
\end{array}
\end{footnotesize} \mbox{and}~
\begin{footnotesize}
\cF_X=
\begin{array}{cccc}
\bullet & \bullet & \bullet & \bullet \\
&\bullet & \bullet & \bullet  \\
&\bullet & \bullet & \bullet
\end{array},
\end{footnotesize}
~\mbox{respectively}.
\end{align*}
\end{example}

\vspace{0.2cm}


Let $| \cF |$ denote the {\it size} of a Ferrers diagram $\cF$,
i.e., the number of dots in $\cF$. A Ferrers diagram of a
$k$-dimensional subspace has size at most $k \cdot (n-k)$. It can
be embedded in a $k\times (n-k)$ box. Let $p(k,\eta,m)$ be the
number of partitions of $m$ whose Ferrers diagram can be embedded
into a box of size $k\times \eta$. The following result was given
in~\cite[pp. 33-34]{And84}.
\begin{lemma}
\label{lem: recursion} $p(k,\eta,m)$ satisfies the following
recurrence relation:
\begin{equation}
\label{eq:rec_part} p(k,\eta,m)=p(k, \eta-1,m-k)+p(k-1,\eta,m)
\end{equation} with the
initial conditions
\begin{equation*}
\label{eq:initial_cond} p(k,\eta,m)=0 ~~ \textrm{if} ~~ m < 0~
\textrm{or} ~ m > \eta \cdot k ~\textrm{and}~ p(k,\eta,0)=1.
\end{equation*}
\end{lemma}
\vspace{0.2cm}

Let $\cF$ be a Ferrers diagram of size $m$ embedded in a $k\times
(n-k)$ box. The number of $k$-dimensional subspaces whose Ferrers
diagram is $\cF$, is equal to $q^m$. By (\ref{eq:num_subspace})
this implies the following theorem~\cite[p. 327]{vLWi92} which
shows the connection between the $q$-ary Gaussian coefficients and
partitions.
\begin{theorem}
\label{thm:vanLint} For any given integers $k$ and $n$, $0 < k
\leq n$, \[ \left[\begin{array}{c}
n\\
k\end{array}\right]_{q}=\sum_{m=0}^{k(n-k)}\alpha_m q^m,\] where
$\alpha_m=p(k,n-k,m)$.
\end{theorem}

\vspace{0.1cm}

The order defined in Section~\ref{sec:Ferrers} is based on
Theorem~\ref{thm:vanLint}. We order the subspaces by the size of
their Ferrers diagrams. The order of Ferrers diagrams with the
same size is explained in Section~\ref{sec:Ferrers}. Two subspaces
with the same Ferrers diagrams are ordered lexicographically by
their Ferrers tableaux forms. This order seems to be the most
natural order of $\Gr$. But, a less natural representation, which
follows, and its related order, will lead to a more efficient
enumerative coding.

Each $k$-dimensional subspace $X \in \Gr$ has an {\it identifying
vector} $v(X)$~\cite{EtSi09}. $v(X)$ is a binary vector of length
$n$ and weight $k$, where the {\it ones} in $v(X)$ are exactly in
the positions (columns) where $\mbox{RE}(X)$ has the leading
coefficients (of the rows).


Let $X\in\Gr$ be a $k$-dimensional subspace. The {\it extended
representation}, $\mbox{EXT}(X)$, of $X$ is a $(k+1)\times n$
matrix obtained by combining the identifying vector
$v(X)=(v(X)_n,\ldots,v(X)_1)$ and the RREF
$\mbox{RE}(X)=(X_n,\ldots,X_1)$, as follows
\begin{align*}
\mbox{EXT}(X)=\left( \begin{array}{cccc}
 v(X)_n & \ldots & v(X)_2 & v(X)_1 \\
 X_n & \ldots & X_2 & X_1
\end{array}
\right).
\end{align*}
Note, that $v(X)_n$ is the most significant bit of $v(X)$. Also,
$X_i$ is a column vector and $v(X)_i$ is the most significant bit
of the column vector $\begin{footnotesize} \left(
\begin{array}{c} v(X)_i
\\X_i
\end{array}\right) \end{footnotesize}$.

\vspace{0.1cm}

\begin{example} Consider the 3-dimensional subspace $X$ of
Example~\ref{exm:running}. Its identifying vector is
$v(X)=1011000$ and its extended representation is given by

\vspace{-0.3cm}

\begin{footnotesize}
\begin{align*}
\mbox{EXT}(X)= \left( \begin{array}{ccccccc}
1 & 0 & 1 & 1 & 0 & 0 & 0\\
1 & 0 & 0 & 0 & 1 & 1 & 0 \\
0 & 0 & 1 & 0 & 1 & 0 & 1 \\
0 & 0 & 0 & 1 & 0 & 1 & 1
\end{array}
\right) ~.
\end{align*}
\end{footnotesize}
\end{example}

\vspace{0.1cm}

The extended representation is redundant since the RREF define a
unique subspace. Nevertheless, this representation will lead to
more efficient enumerative coding. Some insight for this will be
the following well known equality given in~\cite[p. 329]{vLWi92}.
\begin{lemma}
\label{lem:pascal} For all integers $q$, $k$, and $n$, such that
$k\leq n$ we have
\begin{equation}
\label{eq:pascal} \begin{small}\left[\begin{array}{c}
n\\
k\end{array}\right]\end{small}_{q}=q^k\begin{small}\left[\begin{array}{c}
n-1\\
k\end{array}\right]\end{small}_{q}+\begin{small}\left[\begin{array}{c}
n-1\\
k-1\end{array}\right]\end{small}_{q}~.
\end{equation}
\end{lemma}
\vspace{0.4cm} The order defined in Section~\ref{sec:Cover} is
based on Lemma~\ref{lem:pascal} (applied recursively). Note that
the number of subspaces in which $v(X)_1=1$ is
$\begin{footnotesize} \left[\begin{array}{c}
n-1\\
k-1\end{array}\right]_{q}
\end{footnotesize}$ and the number of subspaces in which $v(X)_1=0$
is $\begin{footnotesize} q^k\left[\begin{array}{c}
n-1\\
k\end{array}\right]_{q}
\end{footnotesize}$.

\begin{remark}
A simple connection between (\ref{eq:rec_part}) and
(\ref{eq:pascal}) was given in~\cite[p. 68]{AnEr04}.
\end{remark}

\section{Coding Based on Extended Representation}
\label{sec:Cover}

In this section we define a lexicographic order for the
Grassmannian based on the extended representation. We present an
enumerative coding technique for the Grassmannian using this order
and discuss its complexity.

\subsection{Order for $\Gr$ Based on the Extended Representation}
\label{ssec:lex_order} Let $\{ x \}$ denote the value of
$x=(x_1,x_2,...,x_r) \in \Z_q^r$ (or $x=(x_1,x_2,...,x_r)^T \in
\Z_q^r$), where the vector $x$ is viewed as a number in base-$q$
notation. Let $\{i\}_{q}$ be the base-$q$ representation of the
nonnegative integer $i$. The resulting vector is either a row
vector or a column vector depending on the context.

Let $X,Y\in \Gr$ be two $k$-dimensional subspaces and
$\mbox{EXT}(X)$, $\mbox{EXT}(Y)$ be the extended representations
of $X$ and $Y$, respectively. Let $i$ be the least index such that
$\mbox{EXT}(X)$ and $\mbox{EXT}(Y)$ have different columns. We say
that $X<Y$ if $\begin{footnotesize}\left\{
\begin{array}{c} v(X)_i \\X_i
\end{array}\right\}<\left\{ \begin{array}{c} v(Y)_i \\Y_i
\end{array}\right\}.\end{footnotesize}$ Clearly,
this definition induces an order for $\Gr$.

\vspace{0.1cm}

\begin{example}
For $X,Y,Z\in\mathcal G_2(6,3)$ whose $\mbox{EXT}(X),$
$\mbox{EXT}(Y)$ and $\mbox{EXT}(Z)$ are given by

\begin{footnotesize}
\begin{align*}
\mbox{EXT}(X)=\left(\begin{array}{cccccc}
1 & 1 & 1 & 0 & 0 & 0 \\
1 & 0 & 0 & 0 & 1 & 0 \\
0 & 1 & 0 & 0 & 0 & 0 \\
0 & 0 & 1 & 1 & 0 & 0 \\
\end{array}
\right),
\end{align*}
\end{footnotesize}
\begin{footnotesize}
\begin{align*}\mbox{EXT}(Y)=\left( \begin{array}{cccccc}
1 & 1 & 0 & 1 & 0 & 0 \\
1 & 0 & 0 & 0 & 0 & 0 \\
0 & 1 & 1 & 0 & 0 & 0 \\
0 & 0 & 0 & 1 & 0 & 0
\end{array}
\right),
\end{align*}
\end{footnotesize}
\begin{footnotesize}
\begin{align*}
\mbox{EXT}(Z)=\left( \begin{array}{cccccc}
1 & 1 & 0 & 0 & 1 & 0 \\
1 & 0 & 0 & 0 & 0 & 0 \\
0 & 1 & 0 & 0 & 0 & 0 \\
0 & 0 & 0 & 0 & 1 & 0
\end{array}
\right) ,
\end{align*}
\end{footnotesize}
we have $Y<X<Z$.
\end{example}

\subsection{Enumerative Coding Based on Extended Representation}
\label{ssec:enumaration}

Let $\begin{footnotesize} N\left( \begin{array}{ccc}
 v_j & \ldots &  v_1 \\
 X_j & \ldots &  X_1
\end{array}
\right)
\end{footnotesize}$
be the number of elements in $\Gr$  for which the  first $j$
columns in the extended representation are given by
$\begin{footnotesize} \left( \begin{array}{ccc}
 v_j & \ldots &  v_1 \\
 X_j & \ldots &  X_1
\end{array}
\right).
\end{footnotesize}$

\begin{remark} We view all the $q$-ary vectors of length ${k+1}$
as our finite alphabet. Let $S$ be the set of all $q$-ary
${(k+1)\times n}$ matrices which form extended representations of
some $k$-dimensional subspaces. Now, we can use Cover's method to
encode/decode the Grassmannian. In this setting note that
$\begin{footnotesize} N\left( \begin{array}{ccc}
 v_j & \ldots &  v_1 \\
 X_j & \ldots &  X_1
\end{array}
\right)
\end{footnotesize}$ is equivalent to
$n_S(x_1,x_2,\ldots,x_j)$, where $\left( \begin{array}{c}
 v_i \\
 X_i
\end{array}
\right)$ has the role of $x_i$.
\end{remark}

Let $w_j$ denotes the weight of the first $j$ entries of $v(X)$,
i.e., $w_j=\sum_{\ell=1}^j v_\ell$.
\begin{lemma}
\label{lem:N} For $1 \leq j \leq n$ we have
\[
N\left( \begin{array}{ccc}
 v_j & \ldots &  v_1 \\
 X_j & \ldots &  X_1
\end{array}\right)=\begin{small}\left[\begin{array}{c}
n-j\\k- w_j \end{array}\right]_{q}\end{small}.
\]

\end{lemma}
\vspace{0.01cm}
\begin{proof}
Let $X$ be a $k$-dimensional subspace in $\Gr$  for which the
first $j$ columns in the extended representation are given by
$\begin{footnotesize} \left( \begin{array}{ccc}
 v_j & \ldots &  v_1 \\
 X_j & \ldots &  X_1
\end{array}
\right).
\end{footnotesize} $
Then in the last $n-j$ entries of $v(X)$ there are $k-w_j$
\textit{ones}, and the $w_j$ last rows of $n-j$ last columns of
$\mbox{EXT}(X)$ have only \textit{zeroes}. Therefore, restriction
of $\mbox{EXT}(X)$ to the first $(k+1)-w_j$ rows of the last $n-j$
columns defines a subspace in $\mathcal G_q(n-j,k- w_j ).$ Hence,
we have
\[
N\left( \begin{array}{ccc}
 v_j & \ldots &  v_1 \\
 X_j & \ldots &  X_1
\end{array}\right)=\begin{small}\left[\begin{array}{c}
n-j\\k-w_j\end{array}\right]_{q}\end{small}~.
\]
\end{proof}

\begin{theorem}
\label{thm:Ind} Let $X\in\mathcal{G}_{q}(n,k)$ be a subspace,
where
\[
\mbox{EXT}(X)=\left( \begin{array}{cccc}
 v_n & \ldots & v_2 & v_1 \\
 X_n & \ldots & X_2 & X_1
\end{array}
\right).
\]
Then the lexicographic index (decoding) of $X$,
$\mbox{I}_{\begin{tiny}\mbox{EXT}\end{tiny}}(X)$, is given by
\begin{equation}
\label{eq:Ind2}
\mbox{I}_{\begin{tiny}\mbox{EXT}\end{tiny}}(X)=\sum_{j=1}^{n}(v_{j}q^{k-w_{j-1}}+(1-v_{j})
\frac{\left\{ X_{j}\right\} }{q^{w_{j-1}}})
\begin{footnotesize}\left[\begin{array}{c}n-j\\
k-w_{j-1}\end{array}\right]_{q}\end{footnotesize}
\end{equation}
\end{theorem}

\begin{proof}
By (\ref{cover}) we have that
$\mbox{I}_{\begin{tiny}\mbox{EXT}\end{tiny}}(X)$ is equal to
\begin{equation} \sum_{j=1}^{n}\sum_{\tiny\tiny
\left( \begin{array}{c} u \\W
\end{array}\right)<\left( \begin{array}{c} v_j \\X_j
\end{array}\right)}N\left( \begin{array}{cccc}
u & v_{j-1} & \ldots &  v_1 \\
W & X_{j-1} & \ldots &  X_1
\end{array}\right).\label{ind_2_proof}
\end{equation}

To compute the $j$th summand of (\ref{ind_2_proof}), we
distinguish between two cases.\\
Case 1: $v_j=1$. It implies that $X_j$ has weight one, and its
bottom $w_{j-1} +1$ entries (as a column vector) are an {\it one}
followed by $w_{j-1}$ {\it zeroes}, i.e., $X_j =
\{q^{w_{j-1}}\}_q$. Hence, $\mbox{EXT}(X)$ has the form
\[ \left( \begin{array}{ccccccc}
v_n &\ldots & v_{j+1} &1 & v_{j-1} & \ldots &  v_1 \\
X_n &\ldots & X_{j+1} & \{q^{w_{j-1}}\}_q & X_{j-1} & \ldots &  X_1\\
\end{array}\right)~.
\]\\
Therefore, a subspace $Y \in \Gr$ is lexicographically preceding
$X$, where $\mbox{EXT}(Y)$ has the same first $j-1$ columns as
$\mbox{EXT}(X)$, if and only if $\mbox{EXT}(Y)$ has the form
$$\left( \begin{array}{ccccccc}
v'_n &\ldots & v'_{j+1} & 0 & v_{j-1} & \ldots &  v_1 \\
Y_n & \ldots & Y_{j+1}& Y_j & X_{j-1} & \ldots &  X_1
\end{array}\right)~.
$$
Note, that $Y_j$ has {\it zeroes} in the last $w_{j-1}$ entries
(since the leading coefficients of the last $w_{j-1}$ rows are
contained in $( X_{j-1} ~ \cdots ~ X_1 )$). The first $k-w_{j-1}$
entries of $Y_j$ can have any values.

Therefore, in this case the $j$th summand of (\ref{ind_2_proof})
is equal to
\[\sum_{s=0}^{q^{k-w_{j-1}}-1}N\left( \begin{array}{cccc}
 0 & v_{j-1} & \ldots &  v_1 \\
 \{s\cdot q^{w_{j-1}}\}_q & X_{j-1} & \ldots &  X_1
\end{array}\right)
\]\\
which is equal by Lemma~\ref{lem:N} to
\begin{equation}q^{k-w_{j-1}}\begin{small}\left[\begin{array}{c}
n-j\\
k-w_{j-1}\end{array}\right]_{q}\end{small}.\label{firstSummand}
\end{equation}\\
Case 2: $v_j=0$. Since $w_{j-1}=\sum_{\ell=1}^{j-1}v_\ell$, it
follows that the last $w_{j-1}$ entries of $X_j$ are {\it zeroes},
i.e., $\{X_j\}$ is a multiple of $q^{w_{j-1}}$. Hence,
$\mbox{EXT}(X)$ has the form
\[ \left( \begin{array}{ccccccc}
v_n &\ldots & v_{j+1} &0 & v_{j-1} & \ldots &  v_1 \\
X_n &\ldots & X_{j+1} & X_j & X_{j-1} & \ldots &  X_1\\
\end{array}\right).
\]\\
Therefore, a subspace $Y \in \Gr$ is lexicographically preceding
$X$, where $\mbox{EXT}(Y)$ has the same first $j-1$ columns as
$\mbox{EXT}(X)$, if and only if $\mbox{EXT}(Y)$ has the form
\[ \left( \begin{array}{ccccccc}
v'_n &\ldots & v'_{j+1} &0 & v_{j-1} & \ldots &  v_1 \\
Y_n &\ldots & Y_{j+1} &\{s\cdot q^{w_{j-1}}\}_q & X_{j-1} & \ldots
&  X_1
\end{array}\right),
\]\\
where $0\leq s\leq \frac {\{X_j\}}{q^{w_j-1}}-1$.

Thus, in this case the $j$th summand of (\ref{ind_2_proof}) is
equal to
\[\sum_{s=0}^{\frac {\{X_j\}}{q^{w_{j-1}}}-1}N\left(
\begin{array}{cccc}
 0 & v_{j-1} & \ldots &  v_1 \\
 \{s\cdot q^{w_{j-1}}\}_q & X_{j-1} & \ldots &  X_1
\end{array}\right),
\]\\
which is equal by Lemma~\ref{lem:N} to
\begin{equation}\frac{\left\{ X_{j}\right\} }{q^{w_{j-1}}}\left[
\begin{array}{c}
n-j\\
k-w_{j-1}\end{array}\right]_{q}.\label{secondSummand}
\end{equation}

\vspace{0.2cm}

Finally, combining equations (\ref{firstSummand}) and
(\ref{secondSummand}) in Case 1 and Case~2 implies equation
(\ref{eq:Ind2}).
\end{proof}
\vspace{0.5cm}
\begin{example}
\label{exm:X_0} Let $X\in \mathcal G_2(6,3)$ be a subspace
represented by
\begin{footnotesize}
\begin{align*}
\mbox{EXT}(X)=\left( \begin{array}{cccccc}
0 & 1 & 0 & 1 & 1 & 0 \\
0 & 1 & 1 & 0 & 0 & 1 \\
0 & 0 & 0 & 1 & 0 & 0 \\
0 & 0 & 0 & 0 & 1 & 1
\end{array}
\right) .
\end{align*}
\end{footnotesize}
By Theorem~\ref{thm:Ind} we have that
$$\mbox{I}_{\begin{tiny}\mbox{EXT}\end{tiny}}(X)=5\cdot\begin{footnotesize}\left[\begin{array}{c}5\\3\end{array}\right]_{2}\end{footnotesize}
+2^3\cdot\begin{footnotesize}\left[\begin{array}{c}4\\3\end{array}\right]_{2}\end{footnotesize}
+2^2\cdot\begin{footnotesize}\left[\begin{array}{c}3\\2\end{array}\right]_{2}\end{footnotesize}
+1\cdot\begin{footnotesize}\left[\begin{array}{c}2\\1\end{array}\right]_{2}\end{footnotesize}$$
$$+2\cdot\begin{footnotesize}\left[\begin{array}{c}1\\1\end{array}\right]_{2}\end{footnotesize}
+0\cdot\begin{footnotesize}\left[\begin{array}{c}0\\0\end{array}\right]_{2}\end{footnotesize}=928.
$$
\end{example}

\vspace{0.2cm}

Now, suppose that an index $0 \leq i <
\begin{footnotesize}\left[\begin{array}{c}n\\k\end{array}\right]_q\end{footnotesize}$
is given. Encoding Algorithm A finds $X \in \Gr$ such that
${\mbox{I}_{\begin{tiny}\mbox{EXT}\end{tiny}}(X)=i}$.

\begin{figure}[hbt]
\centering
\begin{algorithm}
\noindent \textit{Encoding Algorithm A:}

\noindent Set $i_{0}=i$, $w_0=0$.

\noindent For $j=1,2,...,n$ do

\begin{itemize}
\item if $w_{j-1}= k$ then set $v_j=v(X)_j=0$, $w_j=w_{j-1}$,
$X_j{=\{0\}_q}$, and $i_j=i_{j-1}$;

\item otherwise

\begin{itemize}
\item if $i_{j-1}\geq q^{k-w_{j-1}}
\begin{footnotesize}\left[\begin{array}{c}
n-j\\k-w_{j-1}\end{array}\right]_{q}\end{footnotesize}$ then set
$v_j{=v(X)_j=1}$, $w_j=w_{j-1}+1$, $X_{j}=\{q^{w_{j-1}}\}_{q}$,
and $i_{j}=i_{j-1}-q^{k-w_{j-1}}\begin{footnotesize}
\left[\begin{array}{c}
n-j\\k-w_{j-1}\end{array}\right]_{q}\end{footnotesize}$;

\item otherwise let $val=\left\lfloor
i_{j-1}/\begin{footnotesize}\left[\begin{array}{c}
n-j\\k-w_{j-1}\end{array}\right]_{q}\end{footnotesize}\right
\rfloor $ and set $v_j{=v(X)_j=0}$, $w_j=w_{j-1}$, $X_{j}=\left\{
val*q^{w_{j-1}} \right\} _{q}$, and
$i_{j}=i_{j-1}-val*\begin{footnotesize}\left[
\begin{array}{c}
n-j\\k-w_{j-1}\end{array}\right]_{q}\end{footnotesize}.$
\end{itemize}
\end{itemize}

\noindent Form the output \[ \mbox{EXT}(X)=\left(
\begin{array}{cccc}
v_n & \ldots & v_2 & v_1 \\
X_n & \ldots & X_2 & X_1
\end{array}
\right).
\]
\end{algorithm}
\end{figure}

\begin{theorem}
Encoding Algorithm A finds the subspace $X \in \Gr$, such that
$\mbox{I}_{\begin{tiny}\mbox{EXT}\end{tiny}}(X)=i$.
\end{theorem}
\begin{proof}
First we will show that the output of the algorithm is  a
$k$-dimensional subspace. In other words, we will prove that the
weight $w_n$ of identifying vector of the resulting subspace $X$
is equal to $k$. We observe that the first "if" of the algorithm
implies that $w_n\leq k$. Note also that $i_j\geq 0$ for all
$1\leq j\leq n$. Suppose that $w_n=k-t$ for some $t>0.$ Let
$n-k+t\leq j'\leq n$  be the last index  where $v(X)_{j'}=0.$ Then
$w_{j'}=k-t-n+j'=w_{j'-1}.$ According to the algorithm,
${i_{j'-1}< q^{k-w_{j'-1}}{\scriptstyle
\begin{footnotesize}\left[\begin{array}{c}
n-j'\\k-w_{j'-1}\end{array}\right]_{q}\end{footnotesize}}=q^{t+n-j'}
{\scriptstyle \begin{footnotesize}\left[\begin{array}{c}
n-j'\\t+n-j'\end{array}\right]_{q}\end{footnotesize}}=0}$ (since
$t>0$), which contradicts the observation that $i_j\geq 0$ for
each $1\leq j\leq n$.

Let $S_j$ be the $j$th summand of
$\mbox{I}_{\begin{tiny}\mbox{EXT}\end{tiny}}(X),$ given in
(\ref{eq:Ind2}), i.e.,
$\mbox{I}_{\begin{tiny}\mbox{EXT}\end{tiny}}(X)=\sum_{t=1}^nS_t.$
To prove the theorem it is sufficient to show that
$i_j=i-\sum_{t=1}^jS_t$ for all $1\leq j\leq n$ and $i_n=0$. The
proof will be inductive.

By the algorithm, for each coordinate $1\leq j\leq n-k$,
\[ i_{j} = \left\{
\begin{array}{cc}i_{j-1}- q^{k-w_{j-1}}
\begin{small}\left[\begin{array}{c} n-j\\k-w_{j-1}\end{array}\right]_{q}\end{small},
& \textrm{if }v(X)_{j}=1\\
i_{j-1}-\frac{\{X_j\}}{q^{w_{j-1}}}
\begin{small}\left[\begin{array}{c}
n-j\\k-w_{j-1}\end{array}\right]_{q}\end{small},
 & \textrm{if }v(X)_{j}= 0\end{array}\right.\]
Thus, \[ i_{j}=i_{j-1}- v(X)_{j} q^{k-w_{j-1}}
\begin{small}\left[\begin{array}{c} n-j\\k-w_{j-1}\end{array}\right]_{q}\end{small} \]
\begin{equation}
\label{index_induction} -(1-v(X)_{j})\frac{\{X_j\}}{q^{w_{j-1}}}
\begin{small}\left[\begin{array}{c} n-j\\k-w_{j-1}\end{array}\right]_{q}\end{small}
=i_{j-1}-S_{j}
\end{equation}
for all $1\leq j\leq n-k$. Thus, for $j=1$ we have
$i_{1}=i-S_{1}$. We assume that $i_{j}=i-\sum_{t=1}^{j}S_{t}$, for
$j\geq1$.
By (\ref{index_induction}), $i_{j+1}=i_{j}-S_{j+1},$ therefore,
$i_{j+1}=i-\sum_{t=1}^{j}S_{t}-S_{j+1}=i-\sum_{t=1}^{j+1}S_{t}.$

Now, we will show that for all $0\leq j\leq n$, $i_j$ is the
lexicographic index of a subspace in $\mathcal{G}_q(n-j,k-w_j)$
with given $j$ first columns of its representation matrix. It will
complete the proof since $i_n$ is the index of subspace in
$\mathcal{G}_q(0,0)$ and thus it is equal to 0.

It is sufficient to prove that $i_j <
\begin{footnotesize}\left[\begin{array}{c}
n-j\\k-w_{j}\end{array}\right]_{q}\end{footnotesize}$ for all
$0\leq j\leq n$. The proof will be inductive. For $j=0$ we observe
that $i_0=i<\begin{footnotesize}\left[\begin{array}{c}
n\\k\end{array}\right]_{q}\end{footnotesize}$ is given. Assume
that $i_{j-1}<\begin{footnotesize}\left[\begin{array}{c}
n-j+1\\k-w_{j-1}\end{array}\right]_{q}\end{footnotesize}$. We will
show that $i_{j}<\begin{footnotesize}\left[\begin{array}{c}
n-j\\k-w_{j}\end{array}\right]_{q}\end{footnotesize}$. We
distinguish between two cases.\\
Case 1: $i_{j-1}\geq
q^{k-w_{j-1}}\begin{footnotesize}\left[\begin{array}{c}
n-j\\k-w_{j-1}\end{array}\right]_{q}\end{footnotesize}$. Then, by
the algorithm, $v_{j}=1$, $w_{j}=w_{j-1}+1$, and
$i_{j}=i_{j-1}-q^{k-w_{j-1}}\begin{footnotesize}\left[\begin{array}{c}
n-j\\k-w_{j-1}\end{array}\right]_{q}\end{footnotesize}$. By the
assumption, $i_{j} < \begin{footnotesize}\left[\begin{array}{c}
n-j+1\\k-w_{j-1}\end{array}\right]_{q}\end{footnotesize}-q^{k-w_{j-1}}
\begin{footnotesize}\left[\begin{array}{c}
n-j\\k-w_{j-1}\end{array}\right]_{q}\end{footnotesize}$ and thus
by  Lemma~\ref{lem:pascal},  $i_{j}\leq
\begin{footnotesize}\left[\begin{array}{c}
n-j\\k-w_{j-1}-1\end{array}\right]_{q}\end{footnotesize}=
\begin{footnotesize}\left[\begin{array}{c}
n-j\\k-w_{j}\end{array}\right]_{q}\end{footnotesize} $.\\
Case 2: $i_{j-1} <
q^{k-w_{j-1}}\begin{footnotesize}\left[\begin{array}{c}
n-j\\k-w_{j-1}\end{array}\right]_{q}\end{footnotesize}$. Then, by
the algorithm, $v_{j}=0$, $w_{j}=w_{j-1}$, and
$$i_{j}=i_{j-1}-\left\lfloor i_{j-1}/
\begin{footnotesize}\left[\begin{array}{c}
n-j\\k-w_{j-1}\end{array}\right]_{q}\end{footnotesize}\right\rfloor
\begin{footnotesize}\left[\begin{array}{c}
n-j\\k-w_{j-1}\end{array}\right]_{q}\end{footnotesize}$$

$$< (\left\lfloor i_{j-1}/\begin{footnotesize}\left[\begin{array}{c}
n-j\\k-w_{j-1}\end{array}\right]_{q}\end{footnotesize}\right\rfloor+1)
\begin{footnotesize}\left[\begin{array}{c}
n-j\\k-w_{j-1}\end{array}\right]_{q}\end{footnotesize}$$

$$-\left\lfloor i_{j-1}/\begin{footnotesize}\left[\begin{array}{c}
n-j\\k-w_{j-1}\end{array}\right]_{q}\end{footnotesize}\right\rfloor
\begin{footnotesize}\left[\begin{array}{c}
n-j\\k-w_{j-1}\end{array}\right]_{q}\end{footnotesize}=
\begin{footnotesize}\left[\begin{array}{c}
n-j\\k-w_{j-1}\end{array}\right]_{q}\end{footnotesize},
$$
since we can write $\lfloor \frac{a}{b}\rfloor\leq a < (\lfloor
\frac{a}{b}\rfloor+1)b$ for all positive integers $a$  and $b$.
\end{proof}
\vspace{0.1cm}
\begin{example}
Let $q=2$, $n=6$, $k=3$, and $i=928$. By using the Encoding
Algorithm A we will find the subspace $X\in \mathcal G_2(6,3)$
such that $\mbox{I}_{\begin{tiny}\mbox{EXT}\end{tiny}}(X)=i$. We
apply the following steps of the algorithm.

\noindent $j=1$: $i_0=928<
2^3\begin{footnotesize}\left[\begin{array}{c}5\\3\end{array}
\right]_{2}\end{footnotesize}=1240$ and hence $v_1=v(X)_1=0,$
$val=\lfloor 928/155 \rfloor=5,$ $\begin{tiny} X_1=\left(
\begin{array}{c}1 \\ 0 \\ 1 \\ \end{array} \right)
\end{tiny}$, and $i_1=928-5\cdot 155=153$.

\noindent $j=2$: $i_1=153\geq
2^3\begin{footnotesize}\left[\begin{array}{c}4\\3\end{array}
\right]_{2}\end{footnotesize}=120$ and hence ${v_2=v(X)_2=1}$,
$\begin{tiny} X_2=\left(
\begin{array}{c}0 \\ 0 \\ 1 \\ \end{array} \right)
\end{tiny}$, and $i_2=153-120=33$.

\noindent $j=3$: $i_2=33 \geq
2^2\begin{footnotesize}\left[\begin{array}{c}3\\2\end{array}
\right]_{2}\end{footnotesize}=28$ and hence $v_3=v(X)_3=1$,
$\begin{tiny} X_3=\left(
\begin{array}{c}0 \\ 1 \\ 0 \\ \end{array} \right)
\end{tiny}$, and $i_3=33-28=5$.

\noindent $j=4$: $i_3=5<
2^1\begin{footnotesize}\left[\begin{array}{c}2\\1\end{array}
\right]_{2}\end{footnotesize}=6$ and hence $v_4=v(X)_4=0,$
$val=\lfloor 5/3\rfloor=1,$ $\begin{tiny} X_4=\left(
\begin{array}{c}1 \\ 0 \\ 0 \\ \end{array} \right)
\end{tiny}$, and $i_4=5-3=2.$

\noindent $j=5$: $i_4=2\geq
2^1\begin{footnotesize}\left[\begin{array}{c}1\\1\end{array}
\right]_{2}\end{footnotesize}=2$ and hence $v_5=v(X)_5=1$,
$\begin{tiny} X_5=\left(
\begin{array}{c}1 \\ 0 \\ 0 \\ \end{array} \right)
\end{tiny}$, and $i_5=2-2=0$.

\noindent
$j=6$: $w_5=3=k$ and hence $v_6=v(X)_6=0$,\\
$\begin{tiny} X_6=\left(
\begin{array}{c}0 \\ 0 \\ 0 \\ \end{array} \right)
\end{tiny}$, and $i_6=i_5=0$.

\vspace{0.2cm}

Therefore, we obtain a subspace $X\in \mathcal G_2(6,3)$ whose
extended representation is given by

\begin{footnotesize}
\begin{align*}
\mbox{EXT}(X)=\left( \begin{array}{cccccc}
0 & 1 & 0 & 1 & 1 & 0 \\
0 & 1 & 1 & 0 & 0 & 1 \\
0 & 0 & 0 & 1 & 0 & 0 \\
0 & 0 & 0 & 0 & 1 & 1
\end{array}
\right) .
\end{align*}
\end{footnotesize}
\end{example}
\subsection{Complexity}
We consider the complexity of computation of lexicographic index
$\mbox{I}_{\begin{tiny}\mbox{EXT}\end{tiny}}( \cdot )$ in
(\ref{eq:Ind2}). Note that all the integers that we use in the
calculations  are $q$-ary integers. Let $M[a,b]$ denotes the
number of operations for the multiplication of two $q$-ary
integers of length $a$ and $b$. It is known~\cite[p. 634]{Knu99},
that for $a>b,$ $M[a,b]=a\log b\log\log b$.

First, we calculate the length of the $q$-ary integer which
represents the largest Gaussian coefficient in (\ref{eq:Ind2}).
This Gaussian coefficient is
$$\begin{small}\left[\begin{array}{c}n-1\\k\end{array}\right]_{q}\end{small}=
\frac{(q^{n-1}-1)\cdots(q^{n-k}-1)} {(q^k-1)\cdots(q-1)},$$ and
hence this length is less than $k(n-k)$.

\noindent If $w_j=w_{j-1}$ then
\begin{equation}
\label{eq:same} \begin{small}\left[\begin{array}{c}n-j\\k-w_{j-1}
\end{array}\right]_{q}\end{small} = \begin{small}\left[\begin{array}{c}n-(j+1)\\k-w_j\end{array}\right]_{q}\end{small}
\cdot\frac {q^{n-j}-1}{q^{n-k-j+w_j}-1}~.
\end{equation}
If $w_j=w_{j-1}+1$ then
\begin{equation}
\label{eq:samep1}
\begin{small}\left[\begin{array}{c}n-j\\k-w_{j-1}
\end{array}\right]_{q}\end{small} = \begin{small}\left[\begin{array}{c}n-(j+1)\\k-w_j\end{array}\right]_{q}\end{small}
\cdot\frac{q^{n-j}-1} {q^{k-w_j+1}-1}~.
\end{equation}
The Gaussian coefficients in (\ref{eq:Ind2}) can be derived from
the identifying vector. Their computation is done by
(\ref{eq:same}) and (\ref{eq:samep1}). Hence, the complexity for
computation of all the Gaussian coefficients that we need in
(\ref{eq:Ind2}) is $O(nM[k(n-k),n])$.

Since multiplication or division by $q^{i}$ is done by a shift of
$i$ digits, there are $n-k$ indices where $v_j=0$, and the length
of $\{X_j\}$ is $k$, it follows that the complexity of these
operations is $O((n-k)M[k(n-k),k])$. Finally, in (\ref{eq:Ind2})
there are at most $n$ additions of integers whose length is at
most $k(n-k+1)$, and therefore the complexity of these operations
can be omitted.

Hence, the complexity of computation of
$\mbox{I}_{\begin{tiny}\mbox{EXT}\end{tiny}} (\cdot)$ in
(\ref{eq:Ind2}) is $O(nM[k(n-k),n])$, i.e., $O(nk(n-k)\log
n\log\log n)$.

Therefore, we have proved the following theorem:
\begin{theorem}
\label{thm:complexity2} The computation complexity of the
lexicographic index (decoding) in (\ref{eq:Ind2}) is
$O(nk(n-k)\log n\log\log n)$ digits operations.
\end{theorem}

\vspace{0.3cm}

If $k < \log n\log\log n$ then the Gaussian coefficients in
(\ref{eq:Ind2}) can be computed more efficiently. For their
computation we can use Lemma~\ref{lem:pascal}. To compute
$\begin{footnotesize}\left[\begin{array}{c}
n\\
k\end{array}\right]_{q}\end{footnotesize}$ we need to compute
$\begin{footnotesize}\left[\begin{array}{c}
\eta\\
\kappa\end{array}\right]_{q}\end{footnotesize}$ for all $\eta$ and
$\kappa$ such that $0 \leq \kappa \leq k$ and $0 \leq \eta -
\kappa \leq n-k$. It requires at most $k(n-k)$ additions of
integers whose length is at most $k(n-k)$, and a total of at most
$k(n-k)$ shifts. All other computations do not change and can be
omitted from the total complexity. Thus, we have

\begin{theorem}
\label{thm:complexity3} If $\min \{ k,n-k \} < \log n\log\log n$,
then the computation complexity of the lexicographic index in
(\ref{eq:Ind2}) is $O(n^2 \min \{ k,n-k \}^2 )$ digits operations.
\end{theorem}

Finally, in a similar way we can show that the computation
complexity of Encoding Algorithm A is the same as the computation
complexity given for the decoding in Theorem~\ref{thm:complexity2}
and in Theorem~\ref{thm:complexity3}.
\section{Coding Based on Ferrers Tableaux Form}
\label{sec:Ferrers} In this section we present an enumerative
coding for the Grassmannian based on the Ferrers tableaux form
representation of $k$-dimensional subspaces. Note, that even so
this enumerative coding is less efficient, it is more intuitive
and might have its own applications. Lexicodes based on the
related order, were found to be larger than the known
codes~\cite{SiEt10}.


\subsection{Enumerative Coding for Ferrers Diagrams of the Same Size}

Let $\cF$ be a Ferrers diagram  of size $m$ embedded in a $k\times
(n-k)$ box. We represent $\cF$ by an integer vector of length
$n-k$, $(\cF_{n-k},...,\cF_2,\cF_1)$, where $\cF_i$ is equal to
the number of dots in the $i$-th column of $\cF$, $1\leq i\leq
n-k$. Note, that the columns are numbered from right to left and
that $0 \leq \cF_{i+1} \leq \cF_i \leq k$ for all $1 \leq i \leq
n-k-1$. Let $\cF$ and $\widetilde{\cF}$ be two Ferrers diagrams of
the same size. We say that $\cF < \widetilde{\cF}$ if $\cF_i
>\widetilde{\cF}_i$ for the least index $i$ such that  $\cF_i \neq
\widetilde{\cF}_i$, i.e., in the least column where they have a
different number of dots, $\cF$ has more dots than
$\widetilde{\cF}$. This is similar to the lexicographic order
defined in the literature for unrestricted partitions,
e.g.~\cite{NMS71},\cite[pp. 93-98]{Rus}.

Let $N_m(\cF_j,...,\cF_2,\cF_1)$ be the number of Ferrers diagrams
of size $m$ embedded in a $k\times (n-k)$ box, for which the first
$j$ columns are given by $(\cF_j,...,\cF_2,\cF_1)$.


\begin{lemma}
\label{lem:calc_Nm} If $1 \leq j \leq n-k$ and $0 < m \leq k(n-k)$
then
\begin{equation*}
N_m(\cF_j,...,\cF_2,\cF_1)=p(\cF_j,n-k-j,m-\sum_{i=1}^j\cF_i).\label{eq:N_m}
\end{equation*}
\end{lemma}
\begin{proof}
The lemma is an immediate consequence from the fact that $\cF =
(\cF_{n-k},...,\cF_2,\cF_1)$ is a Ferrers diagram with $m$ dots
embedded in a $k \times (n-k)$ box if and only if
$(\cF_{n-k},...,\cF_{j+1})$ is a Ferrers diagram with
$m-\sum_{i=1}^j\cF_i$ dots embedded in an $\cF_j \times (n-k-j)$
box.
\end{proof}
\begin{remark}
We view the set $\Z_{k+1} = \{ 0,1,\ldots , k \}$ as our finite
alphabet since $0 \leq \cF_i \leq k$. Let $S$ be the set of all
$(n-k)$-tuples over $\Z_{k+1}$ which represent Ferrers diagrams
embedded in a $k \times (n-k)$ box. In other words,
${(\cF_{n-k},...,\cF_2,\cF_1) \in S}$ if and only if $0 \leq \cF_i
\leq \cF_{i-1} \leq k$ for each $2 \leq i \leq n-k$. Now, we can
use Cover's method to encode/decode the set of Ferrers diagrams
with $m$ dots embedded in a $k \times (n-k)$ box. In this setting
note that $N_m(\cF_j,...,\cF_2,\cF_1)$ is equivalent to
$n_S(x_1,x_2,\ldots,x_j)$, where $\cF_i$ has the role of $x_i$.
\end{remark}

\begin{theorem}
\label{thm:calc_Nm} Let $\cF = (\cF_{n-k},...,\cF_2,\cF_1)$ be a
Ferrers diagram of size $m$ embedded in a $k \times (n-k)$ box.
Then the lexicographic index (decoding), $\mbox{ind}_m$, of $\cF$
among all the Ferrers diagrams with the same size $m$ is given by
\begin{equation}
\label{eq:ind_m} \mbox{ind}_m(\cF)=\sum_{j=1}^{n-k}
\sum_{a=\cF_j+1}^{\cF_{j-1}}p(a,n-k-j,m-\sum_{i=1}^{j-1}\cF_i -a),
\end{equation}
where we define $\cF_{0}=k.$
\end{theorem}
\begin{proof}
By (\ref{cover}) we have that
\begin{equation*}
\mbox{ind}_m(\cF)=\sum_{j=1}^{n-k}
\sum_{a=\cF_j+1}^{\cF_{j-1}}N_m(a,\cF_{j-1},...,\cF_2,\cF_1).
\end{equation*}
The theorem follows now from Lemma~\ref{lem:calc_Nm}.
\end{proof}

\begin{remark}
The summation in Theorem~\ref{thm:calc_Nm} is over larger values,
while the summation in (\ref{cover}) is over smaller values, due
to the defined order ($\cF < \widetilde{\cF}$ if $\cF_i
>\widetilde{\cF}_i$ for the least index $i$).
\end{remark}

\vspace{0.1cm}

Theorem~\ref{thm:calc_Nm} implies that if we can calculate
$p(k,\eta,m)$ efficiently then we can calculate
$\mbox{ind}_m(\cF)$ efficiently for a Ferrers diagram of size $m$
embedded in a $k\times (n-k)$ box.

Now suppose that an index $0 \leq i < p(k,n-k,m)$ is given.
Encoding Algorithm B finds a Ferrers diagram $\cF$  of size $m$
embedded in a $k\times(n-k)$ box, such that $\mbox{ind}_m(\cF)=i$.

\begin{figure}[hbt]
\centering
\begin{algorithm}
\noindent \textit{Encoding Algorithm B:}

\textit{Step 1}: Set $\cF_0=k ,~ \ell_1=0,~h=i, ~ i_0=i$;

\begin{itemize}
\item while $h\geq N_m(\cF_0-\ell_1)$ set $h=h-N_m(\cF_0-\ell_1)$,
$\ell_1=\ell_1+1$;

\item set $\cF_1=\cF_0-\ell_1$, and $ i_1=h$;
\end{itemize}

\textit{Step 2}: For $j=2,...,n-k$ do
\begin{itemize}
\item if $\sum_{i=1}^{j-1}\cF_i=m$ then set $\cF_j=0$;

\item otherwise do

\noindent begin

\begin{itemize}
\item set $\ell_j=0,h=i_{j-1}$;

\item while $h\geq N_m(\cF_{j-1}-\ell_j,\cF_{j-1},...,\cF_1)$ set
$h=h-N_m(\cF_{j-1}-\ell_j,\cF_{j-1},...,\cF_1)$,
$\ell_j=\ell_{j}+1$;

\item set $\cF_j=\cF_{j-1}-\ell_j$, and $ i_j=h$;
\end{itemize}
\end{itemize}
\indent ~~~~~ end \{begin\}

\textit{Step 3}: Form the output $\cF =
(\cF_{n-k},...,\cF_2,\cF_1)$.

\end{algorithm}
\end{figure}
\begin{remark}
We did not join Step 1 and Step 2, since
$N_m(\cF_{j-1}-\ell_j,\cF_{j-1},...,\cF_1)$ is not defined for
$j=1$.
\end{remark}

\subsection{Order for $\Gr$ Based on Ferrers Tableaux Form}
Let $X,\: Y\in\Gr$ be two $k$-dimensional subspaces and let
$\cF_X$, $\cF_Y$ be the related Ferrers diagrams. Let
$x=(x_1,x_2,...,x_{|\cF_X|})$ and $y=(y_1,y_2,...,y_{|\cF_Y|})$ be
the entries vectors of $\cF(X)$ and $\cF(Y)$, respectively. These
entries are numbered from right to left, and from top to bottom.


We say that $X < Y$ if one of the following conditions holds.
\begin{itemize}
\item $| \cF_X | > | \cF_Y | $;

\item $| \cF_X | = | \cF_Y | $ and  $\cF_X < \cF_Y$;

\item $\cF_X = \cF_Y $ and $\{ x \} < \{ y \}.$
\end{itemize}

\noindent Clearly, this definition induces an order for $\Gr$.

\vspace{0.1cm}
\begin{example}
Let $X,Y,Z,W\in\mathcal G_2(6,3)$ be given by
\begin{footnotesize}
\begin{align*}
\cF(X)=
\begin{array}{ccc}
1 & 1 & 1 \\
1 & 1 & 1  \\
&  & 1
\end{array},~~~
\cF(Y)=
\begin{array}{ccc}
1 & 0 & 1 \\
& 0 & 0  \\
& 1 & 1
\end{array},
\end{align*}
\begin{align*}
\cF(Z)=
\begin{array}{ccc}
1 & 1 & 1 \\
& 1 & 1  \\
& & 0
\end{array},~~~
\cF(W)=
\begin{array}{ccc}
1 & 1 & 1 \\
& 1 & 1  \\
& & 1
\end{array}.
\end{align*}
\end{footnotesize}
$\cF_Z =\cF_W$ and by definition
$Z < W$. Clearly, $|\cF_X | = | \cF_Y | > | \cF_Z |$ and $\cF_Y <
\cF_X$. Thus, $Y < X < Z < W$.
\end{example}

\subsection{Enumerative Coding Based on Ferrers Tableaux Form}
In this subsection, we use the given order of Ferrers tableaux
forms and Theorem~\ref{thm:vanLint} for enumerative coding for
$\Gr$.
\begin{theorem}
\label{thm:Ferrers_index} Let $X \in \Gr$, $\cF_X$ be the Ferrers
diagram of $X$, and let $x= (x_1,x_2,...,x_{|\cF_X|})$ be the
entries vector of $\cF(X)$. Then the lexicographic index
(decoding) of $X$, $\mbox{Ind}_{\cF} (X)$, defined by the order
based on Ferrers tableaux form, is given by
\begin{equation}\mbox{Ind}_{\cF}(X)=\sum_{i=|\cF_X|+1}^{k(n-k)}\alpha_{i}q^{i}+
\mbox{ind}_{|\cF_X|}(\cF_X)q^{|\cF_X|}+\{x\},\label{Ind1}
\end{equation}
where $\alpha _i$, $|\cF_X|+1 \leq i \leq k(n-k)$, is defined in
Theorem~\ref{thm:vanLint}.
\end{theorem}

\begin{proof} To find $\mbox{Ind}_{\cF}(X)$ we have to calculate the number of
$k$-dimensional subspaces which are preceding $X$ according to the
order defined above.

\begin{enumerate}
\item All the $k$-dimensional subspaces with Ferrers diagrams
which have more dots than $\cF_X$ are preceding $X$. Their number
is $\sum_{i=|\cF_X|+1}^{k(n-k)}\alpha_{i}q^{i}$.

\item There are $\mbox{ind}_{|\cF_X|}(\cF_X)$ Ferrers diagrams
with $|\cF_X|$ dots which are preceding $X$. Hence, there are
$\mbox{ind}_{|\cF_X|}(\cF_X)q^{|\cF_X|}$ $k$-dimensional subspaces
whose Ferrers diagrams have $|\cF_X|$ dots and preceding $X$.

\item Finally, the number of $k$-dimensional subspaces whose
Ferrers diagram is $\cF_X$ which are preceding $X$ is $\{x\}$.
\end{enumerate}
\end{proof}

\begin{example}
Let $X\in\mathcal{G}_2(6,3)$ be the subspace of
Example~\ref{exm:X_0}, whose Ferrers tableaux form and Ferrers
diagram are
$$\begin{footnotesize}
\cF(X)=
\begin{array}{cc}
1 & 1  \\
 & 0  \\
 & 1
\end{array}
\end{footnotesize}\;\mbox{and}\;
\begin{footnotesize}
\cF_X=
\begin{array}{cc}
\bullet & \bullet  \\
 & \bullet   \\
 & \bullet
\end{array}.
\end{footnotesize} $$

By Theorem~\ref{thm:Ferrers_index} we have that
$$\mbox{Ind}_\cF(X)=\sum_{i=5}^{9}\alpha_{i}2^{i}+
\text{ind}_{4}(\cF_X)2^4+\{(1011)\}.$$ Since $\alpha_5=3$,
$\alpha_6=3$, $\alpha_7=2$, $\alpha_8=1$, $\alpha_9=1$
(see~\cite[pp. 326-328]{vLWi92}), $\text{ind}_4(\cF_X)=0$, and
$\{(1011)\}=11$, it follows that $\mbox{Ind}_\cF(X)=1323$.

\end{example}

\vspace{0.15cm}

Now suppose that an index $0 \leq i <
\begin{footnotesize}\left[\begin{array}{c}n\\k\end{array}\right]_{q}\end{footnotesize}$ is given.
Encoding Algorithm C finds a subspace $X \in \Gr$ such that
$\mbox{Ind}_{\cF}(X)=i$.

\vspace{-0.2cm}

\begin{figure}[hbt]
\centering
\begin{algorithm}
\noindent \textit{Encoding Algorithm C:}

\noindent Set $i_0=i$.

\noindent For $j=0,\ldots, k(n-k)$ do

\begin{itemize}
\item if $i_j< \alpha_{k(n-k)-j}q^{k(n-k)-j}$ then set
$|\cF_X|{=k(n-k)-j}$,
$\cF_X=\mbox{ind}_{|\cF_X|}^{-1}(\lfloor\frac{i_j}{q^{k(n-k)-j}}\rfloor)$;
$\{i_{j}{-\lfloor\frac{i_j}{q^{k(n-k)-j}}\rfloor
q^{k(n-k)-j}\}_q}$ is assigned to $x$ (the entries vector of
$\cF(X)$) and stop;

\item otherwise set $i_{j+1}=i_j-\alpha_{k(n-k)-j}q^{k(n-k)-j}$.
\end{itemize}

\end{algorithm}
\end{figure}
\subsection{Complexity}
We consider the complexity of the calculation of the lexicographic
index $\mbox{Ind}_{\cF} (X)$, for $X \in \Gr$, whose Ferrers
diagram is $\cF_X = (\cF_{n-k},...,\cF_2,\cF_1)$. We will use the
following lemma concerning partitions to find a bound on the
length of $q$-ary integers which represent the value of
$p(k,n-k,i)$.

\begin{lemma}
\label{lem:bound_p} For any given $n$, $k$, and $i$, we have
${p(k,n-k,i)< e^{\pi \sqrt{\frac{2}{3}i}}}$.
\end{lemma}
\begin{proof}
Clearly, $p(k,n-k,i)\leq p(i)$, where $p(i)$ is the number of
unrestricted partitions of $i$. It is known~\cite[p. 160]{vLWi92}
that $p(i)< e^{\pi \sqrt{\frac{2}{3}i}}$ and the lemma follows.
\end{proof}
\begin{theorem}
\label{thm:complexity1} The computation complexity of the
lexicographic index (decoding) in (\ref{eq:Ind_1_p}) is $O(k^{5/2}
(n-k)^{5/2})$ digit operations.
\end{theorem}
\begin{proof}
First, we combine the expressions in (\ref{eq:ind_m}) and
(\ref{Ind1}) to obtain:
\begin{equation*}
\label{eq:Ind_1_p}
\mbox{Ind}_{\cF}(X)=\sum_{i=|\cF_X|+1}^{k(n-k)}p(k,n-k,i)q^{i}+\{x\}
\end{equation*}
\begin{equation}
\label{eq:Ind_1_p} +q^{|\cF_X|}\sum_{j=1}^{n-k}
\sum_{a=\cF_j+1}^{\cF_{j-1}}p(a,n-k-j,|\cF_X|-
\sum_{i=1}^{j-1}\cF_i-a).
\end{equation}
By the recurrence relation of Lemma~\ref{lem: recursion}, we can
compute the table of $p(j,\ell,i)$ for $j\leq k$, $\ell \leq
\eta$, and $i\leq m$ with no more than $mk\eta$ additions. By
Lemma~\ref{lem:bound_p} each integer in such addition has $O(
\sqrt{k(n-k)})$ digits. Therefore, the computation of all the
values which are needed from the table takes
${O(k^{5/2}(n-k)^{5/2})}$ digit operations.

The number of additions in~(\ref{eq:Ind_1_p}) is $O(k(n-k))$. Each
integer in this addition has $O(k(n-k))$ digits (as a consequence
of Lemma~\ref{lem:bound_p} and the powers of $q$
in~(\ref{eq:Ind_1_p})). The multiplication by $q^i$ is a shift by
$i$ symbols. Hence, these additions and shifts do not increase the
complexity.
\end{proof}

Similarly, we can prove the following theorem.
\begin{theorem}
\label{thm:complexityD1} The computation complexity of Encoding
Algorithm C is ${O(k^{5/2} (n-k)^{5/2})}$ digit operations.
\end{theorem}

\begin{remark}
\label{rem:large} If $k(n-k)-|\cF_X|$ is a small integer then the
complexity of the computation becomes much smaller than the
complexity given in Theorems~\ref{thm:complexity1}
and~\ref{thm:complexityD1}. For example, if $|\cF_X|=k(n-k)$ then
the complexity of the enumerative decoding is $O(k(n-k))$ since
$\mbox{Ind}_{\cF}(X)=\{x\}$ in~(\ref{eq:Ind_1_p}).
\end{remark}

It is worth to mention in this context that the number of
operations in the algorithms can be made smaller if we will
consider the following two observations~\cite[p. 47]{And84}:

\begin{itemize}
\item If $m_1 < m_2 \leq \frac{k \eta}{2}$ then $p(k, \eta ,m_1)
\leq p(k, \eta ,m_2)$.

\item $p(k,\eta,m)=p(k,\eta,k\eta-m)$ and hence we can assume that
$m\leq \frac{k \eta }{2}$.
\end{itemize}

\section{Combination of the Coding Techniques}
\label{sec:combination}

By Theorems~\ref{thm:complexity2},~\ref{thm:complexity3},
and~\ref{thm:complexity1}, it is clear that the enumerative coding
based on the extended representation is more efficient than the
one based on Ferrers tableaux form. But, for some of
$k-$dimensional subspaces of $\F_q^n$ the enumerative coding based
on Ferrers tableaux form is more efficient than the one based on
the extended representation (see Remark~\ref{rem:large}). This is
the motivation for combining the two methods.

The only disadvantage of the Ferrers tableaux form coding is the
computation of the $\alpha_i$'s and $\mbox{ind}_{|\cF_X|}(\cF_X)$
in Theorem~\ref{thm:Ferrers_index}. This is the reason for its
relatively higher complexity. The advantage of this coding is that
once the values of the $\alpha_i$'s and the value of
$\mbox{ind}_{|\cF_X|}(\cF_X)$ are known, the computation of
$\mbox{Ind}_{\cF} (X)$, for $X \in \Gr$, is immediate. Our
solutions for the computation of the $\alpha_i$'s and
$\mbox{ind}_{|\cF_X|}(\cF_X)$ are relatively not efficient and
this is the main reason why we suggested to use the enumerative
coding based of the RREF and the identifying vector of a subspace.
The only disadvantage of this enumerative coding is the
computation of the Gaussian coefficients in (\ref{eq:Ind2}). It
appears that a combination of the two methods is more efficient
than the efficiency of each one separately. The complexity will
remain $O(nk(n-k)\log n\log\log n)$, but the constant will be
considerably reduced on the average. This can be done if there
won't be any need for the computation of the $\alpha_i$'s and the
computation of $\mbox{ind}_{|\cF_X|}(\cF_X)$ will be efficient.

It was proved in~\cite{KK} that $q^{k(n-k)} <
\begin{footnotesize}\left[\begin{array}{c}n\\k\end{array}\right]_q\end{footnotesize} < 4 q^{k(n-k)}$
for $0 < k < n$. Thus, more than $\frac{1}{4}$ of the
$k$-dimensional subspaces in $\Gr$ have the unique Ferrers diagram
with $k(n-k)$ dots, where the identifying vector consists of $k$
{\it ones} followed by $n-k$ {\it zeroes}. All the codewords of
the Reed-Solomon-like code in~\cite{KK} have this Ferrers diagram.
Note that most of the $k$-dimensional subspaces have Ferrers
diagrams with a large number of dots. We will encode/decode these
subspaces by the Ferrers tableaux form coding and the other
subspaces by the extended representation coding. We will choose a
set $S_{\cF}$ with a small number of Ferrers diagrams. $S_{\cF}$
will contain the largest Ferrers diagrams. The Ferrers tableaux
form coding will be applied on these diagrams.


We say that a subspace $X \in \Gr$ is of Type $S_{\cF}$ if $\cF_X
\in S_{\cF}$. In the new order these subspaces are ordered first,
and their internal order is defined as the order of the Ferrers
tableaux forms in Section~\ref{sec:Ferrers}. The order of the
other subspaces is defined by the order of the extended
representation in Section~\ref{sec:Cover}. We define a new index
function
$\mbox{I}_{\begin{scriptsize}\mbox{comb}\end{scriptsize}}$ as
follows:
\begin{equation}
\label{eq:combination}
\mbox{I}_{\begin{scriptsize}\mbox{comb}\end{scriptsize}}(X)=
\left\{
\begin{array}{cc}
\mbox{Ind}_{\cF}(X) & \cF_X\in S_{\cF}\\
\mbox{I}_{\begin{tiny}\mbox{EXT}\end{tiny}}(X)+\Delta_X (S_{\cF})
& \textrm{otherwise}\end{array},\right.
\end{equation}
where $\Delta_X(S_{\cF})$ is the number of subspaces of Type
$S_{\cF}$, which are lexicographically succeeding $X$ by the
extended representation ordering. These $\Delta_X(S_{\cF})$
subspaces are preceding $X$ in the ordering induced by combining
the two coding methods.

We demonstrate the method for the simple case where $S_{\cF}$
consists of the unique Ferrers diagram with $k(n-k)$ dots.

\begin{lemma}
\label{lem:Delta} Let $S_{\cF}$ be a set of Ferrers diagrams,
embedded in a $k \times (n-k)$ box, which contains only one
Ferrers diagram, the unique one with $k(n-k)$ dots. Let $X \in
\Gr$, $X \not\in S_{\cF}$, $\mbox{RE}(X)=(X_n,\ldots,X_1)$, and
let $\ell$, $0\leq \ell \leq n-k-1$, be the number of consecutive
\textit{zeroes} before the first \textit{one} (from the right) in
the identifying vector $v(X)$. Then $\Delta_X(S_{\cF})=
\sum_{i=1}^{\ell}(q^k-1-\{X_i\})q^{k(n-k-i)}$.
\end{lemma}

\begin{proof} If $\ell=0$ then $v(X)_1=1$ and hence
there are no subspaces of Type $S_{\cF}$ which are
lexicographically succeeding $X$ and hence $\Delta_X(S_{\cF})=0$.
For $1\leq \ell \leq n-k-1,$ let $X_1,...,X_\ell$ be the first
$\ell$ columns of $\mbox{RE}(X)$. All the subspaces of Type
$S_{\cF}$ in which the value of the first column is greater than
$\{ X_1 \}$, are lexicographically succeeding $X$. There are
$(q^k-1-\{X_1\})q^{k(n-k-1)}$ such subspaces. All the subspaces of
Type $S_{\cF}$ in which the first $i-1$ columns, $2\leq i\leq
n-k-1$, are equal to the first $i-1$ columns of $\mbox{RE}(X)$,
and the value of the $i$th column is greater than $\{X_i\}$, are
lexicographically succeeding $X$. There are
$(q^k-1-\{X_{i}\})q^{k(n-k-i)}$ such subspaces. Therefore, there
are $\sum_{i=1}^{\ell}(q^k-1-\{X_i\})q^{k(n-k-i)}$ subspaces of
Type $S_{\cF}$ which are lexicographically succeeding $X$ by the
extended representation ordering.
\end{proof}

\begin{example}
Let $X$ be the subspace of Example~\ref{exm:X_0}. By
Example~\ref{exm:X_0} we have
$\mbox{I}_{\begin{tiny}\mbox{EXT}\end{tiny}}(X)=928$, and by
Lemma~\ref{lem:Delta} we have
$\Delta_X(S_{\cF})=(2^3-1-5)2^{3\cdot 2}=2^7$. Hence,
$\mbox{I}_{\begin{scriptsize}\mbox{comb}\end{scriptsize}}(X)=\mbox{I}_{\begin{tiny}\mbox{EXT}\end{tiny}}(X)+\Delta_X(S_{\cF})=928+128=1056$.
\end{example}
\vspace{0.3cm}

Now, suppose that an index $0 \leq i <
\begin{footnotesize}\left[\begin{array}{c}n\\k\end{array}\right]_q\end{footnotesize}$ is given. Based on
(\ref{eq:combination}) and Lemma~\ref{lem:Delta} we can find the
subspace $X$ such that
$\mbox{I}_{\begin{scriptsize}\mbox{comb}\end{scriptsize}} (X)=i$,
where $S_{\cF}$ consists of the unique Ferrers diagram with
$k(n-k)$ dots. We omit the details of the encoding algorithm.


\section{Conclusion}
\label{sec:conclude}

Three methods of enumerative coding for the Grassmannian are
presented. The first is based on the representation of subspaces
by their identifying vector and their reduced row echelon form.
The second is based on the Ferrers tableaux form representation of
subspaces. The complexity of the first method is superior on the
complexity of the second one. The third method is a combination of
the first two. On average it reduces the constant in the first
term of the complexity compared to the complexity of the first
method. Improving on these methods is a problem for future
research.

The enumerative coding is based on an order for the Grassmannian
related to a specific representation. This order can be used to
form lexicographic codes~\cite{CoSl86} in the Grassmannian. To our
surprise some of these lexicographic codes form the best known
error-correcting codes in the Grassmannian. For example, a
lexicode of size 4605 in $\mathcal G_2(8,4)$ with minimum subspace
distance 4 (see~\cite{KK} for the distance definition) was
generated based on Ferrers tableaux form order (compared to the
largest previously known code of size 4573 generated by a
multilevel construction~\cite{EtSi09}). These codes also revealed
a new method to form error-correcting codes in the Grassmannian.
This topic is considered in~\cite{SiEt10}.

Construction of a lexicode might require to generate all subspaces
of $\Gr$ by the given lexicographic order. Usually, this does not
require to use the enumerative coding since the subspaces are
generated one after another. By using one of our orders it is not
difficult to prove that given a subspace $X \in \Gr$, it takes no
more than $O(k n)$ digit operations to generate the next subspace.

\section*{Acknowledgment}

We thank the anonymous reviewers whose comments have helped to
improve the presentation of this paper.



\newpage
\textbf{Natalia Silberstein} was born in Novosibirsk, Russia, in
1977. She received the B.A. and M.Sc. degrees from the Technion -
Israel Institute of Technology, Haifa, Israel, in 2004 and 2007,
respectively, from the Computer Science Department and the Applied
Mathematics Department, respectively. She is currently working
toward the Ph.D. degree in the department of Computer Science  at
the Technion. Her research interests include algebraic
error-correction coding, coding theory, and combinatorial designs.

\vspace{1cm}

{\bf Tuvi Etzion} (M'89-SM'94-F'04) was born in Tel Aviv, Israel,
in 1956. He received the B.A., M.Sc., and D.Sc. degrees from the
Technion - Israel Institute of Technology, Haifa, Israel, in 1980,
1982, and 1984, respectively.

From 1984 he held a position in the department of Computer Science
at the Technion, where he has a Professor position. During the
years 1986-1987 he was Visiting Research Professor with the
Department of Electrical Engineering - Systems at the University
of Southern California, Los Angeles. During the summers of 1990
and 1991 he was visiting Bellcore in Morristown, New Jersey.
During the years 1994-1996 he was a Visiting Research Fellow in
the Computer Science Department at Royal Holloway College, Egham,
England. He also had several visits to the Coordinated Science
Laboratory at University of Illinois in Urbana-Champaign during
the years 1995-1998, two visits to HP Bristol during the summers
of 1996, 2000,  a few visits to the department of Electrical
Engineering, University of California at San Diego during the
years 2000-2010, and several visits to the Mathematics department
at Royal Holloway College, Egham, England, during the years
2007-2009.

His research interests include applications of discrete
mathematics to problems in computer science and information
theory, coding theory, and combinatorial designs.

Dr Etzion was an Associate Editor for Coding Theory for the IEEE
Transactions on Information Theory from 2006 till 2009.

\end{document}